\documentclass[preprint,showpacs,preprintnumbers,amsmath,amssymb]{revtex4}

\usepackage{graphicx}

\begin{document}


\title{\emph{Ab initio} Study of Diketo-Pyrrolo-Pyrrole Polymers for Photovoltaic Applications}

\author{Simon L\'evesque}%
\email{s.levesque@umontreal.ca}%
\author{Paolo E. Trevisanutto}%
\email{paolo.emilio.trevisanutto@umontreal.ca}%
\author{Jean Fr\'ed\'eric Laprade}%
\author{Michel C\^ot\'e}%
\email{michel.cote@umontreal.ca}%
\affiliation{%
D\'epartement de physique et Regroupement qu\'eb\'ecois sur les\\ mat\'eriaux de pointe (RQMP), Universit\'e de Montr\'eal, Canada
}%


\date{\today}

\begin{abstract}
Using density-functional theory with the hybrid functional B3LYP, we investigate the electronic properties of polymers with diketo-pyrrolo-pyrrole (DPP) unit. 
We note that some of the polymers studied have LUCO energy similar to the C70-PCBM, or even lower, making them promising candidates for electron transport in organic photovoltaic devices. 
The homopolymer of DPP is predicted to have a band gap around 1.2 eV and shows a good dispersion of the conduction band. 

\end{abstract}

\pacs{Valid PACS appear here}
\maketitle
\begin{figure}[!t]
\includegraphics[scale=.39]{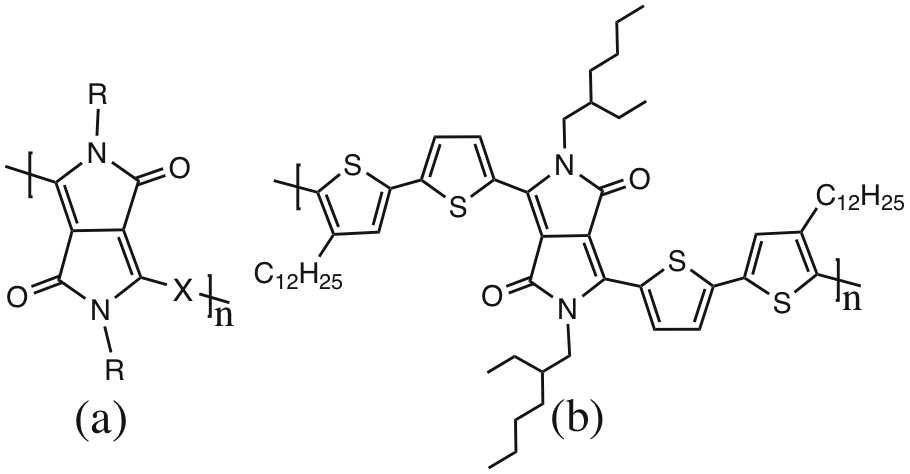}
\caption{(a) Structure of the DPP polymers used in this study and (b) pBBTDPP2 synthetized polymer. The side chains (R) have been remplaced by CH$_3$ for the calculations reported in Fig. \ref{fig2} and by an H for those reported in Fig. \ref{fig3}.}
\label{fig1}
\end{figure}

Organic solar cells are the subject of intense research at this time due to their low fabrication cost, light weight, mechanical flexibility and possible transparency. %
The primary challenge is to raise their power conversion efficiencies (PCEs) from the current maximum achieved of 6\%,\cite{PAR09} to at least 10\% so that they can become economically viable and enable widespread uses. %
In order to improve the efficiency in bulk heterojunction (BHJ) or multilayer organic solar cells, it seems crucial to find polymers with high carrier mobility.\cite{Str08} %
The PCE is also related to the magnitude of the open-circuit voltage ($V_{oc}$),\cite{Sho61} which is linked to the energy difference between the highest occupied crystal orbital (HOCO) level of the donor and the lowest unoccupied crystal orbital (LUCO) level of the acceptor molecules.\cite{Gad04, Bru08, Koo07} %
In their paper, Scharber \emph{et al.},\cite{Sch06} used a simple model to link the PCE to the band gap and the LUCO level of the donor in devices which uses [6,6]-phenyl-C61-butyric acid methyl ester (PCBM) as an acceptor and demonstrated the 10\% PCE was possible with polymers that have the right electronic properties. %
Any hints from theoretical calculations on how one can experimentally tune these levels is then of considerable interest in the search of more efficient solar cells. %
Among the promising polymers, those made with diketo-pyrrolo-pyrrole (DPP) get a lot of attention presently because they have been shown to have relatively high mobilities and PCEs.\cite{Bur08, Wie08, Pee09} %
In this article, we report the results of calculations using density-functional theory (DFT)~\cite{Hoh64, Koh65} on polymers made with the DPP unit.  The general structure of the polymers studied is shown in Fig.\ref{fig1}a, where the side chains are denoted by R and the variable component of the unit cell is denoted by X. %
Fig.\ref{fig1}b shows a polymer already synthetized, poly[3,6-bis-(4'-dodecyl-[2,2']bithiophenyl-5-yl)-2,5-bis-(2-ethyl-hexyl)-2,5-dihydropyrrolo[3,4-]pyrrole-1,4-dione] (pBBTDPP2), which has been used along with PCBM in solar cell reaching 4\% of PCE.\cite{Wie08} %

\begin{figure*}[!t]
\includegraphics[scale=.5]{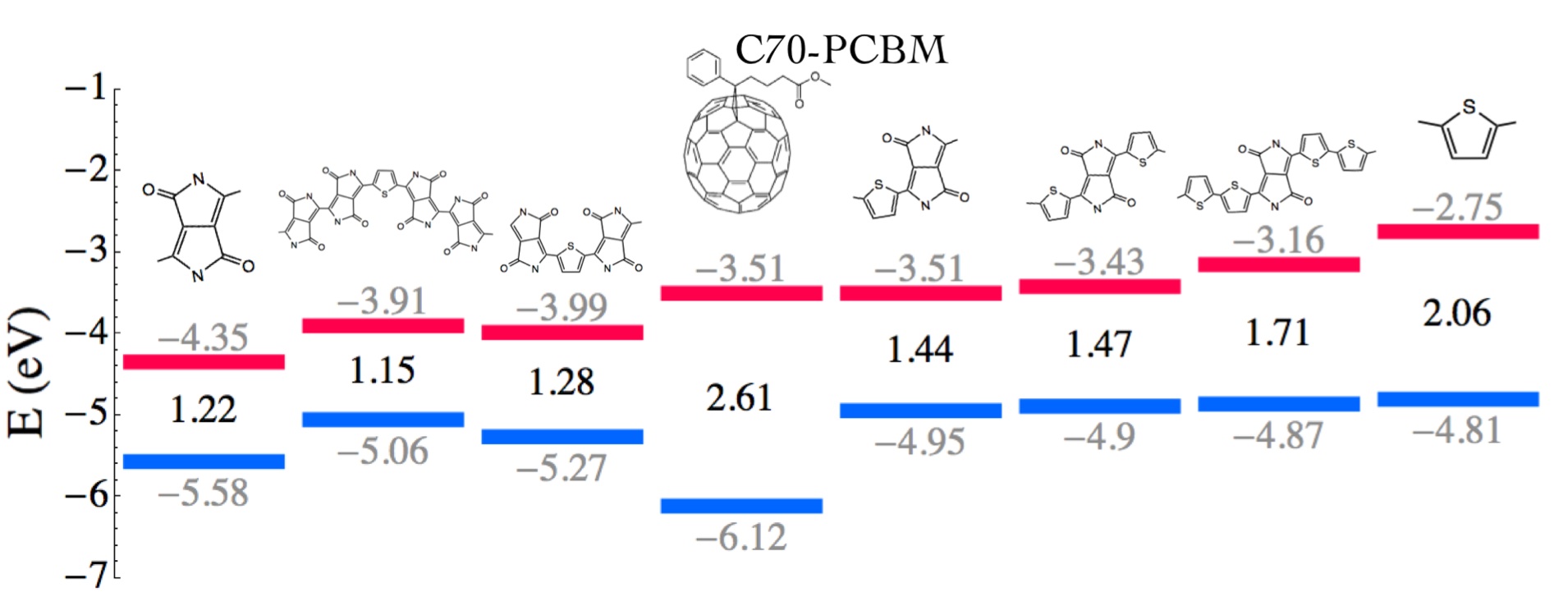}
\caption{(Color online) DFT-B3LYP energies of DPP-type polymers as a function of the number of thiophene unit in the copolymer. LUCO levels, energy gaps and HOCO levels are shown for all polymers.}
\label{fig2}
\end{figure*}

The calculations reported on DPP-type polymers (see Fig~\ref{fig2}) are performed with \texttt{Gaussian~03},\cite{Gau03} %
a code oriented towards molecules using gaussian functions as basis set. The B3LYP %
 functional~\cite{Bec93,Ste94} including exact exchange and the basis 6-311G(d,p) %
 have been used. The geometry of the unit cell and the dihedral angle between two successive units have been optimized for all polymers. We will use the HOCO-LUCO gaps of DFT-B3LYP as an indication of the optical gap of the polymers even if these calculations do not formally include electron-hole (e-h) interaction. 
 To reinforce the DFT conclusions, we use the GW~\cite{HED65} method (GWA) 
to calculate some electronic and optical structures. However, since these calculations involve more intense computation, we could only carry out those for the smaller systems, namely polythiophene (PT) and DPP homopolymer (pDPP). In GWA, the self energy $\Sigma$ 
is approximated to be the convolution of Green's function $G$ and the screened Coulomb potential $W$:
\begin{equation}
\Sigma(\omega)=\int{d\omega' e^{i\omega'0^{+}}G(\omega'+\omega)W(\omega').}
 \label{GW} 
\end{equation}
The $G$ and $W$ functions are calculated using the \textit{ab initio} ground state DFT-local density approximation (LDA) results for infinite isolated polymer chains as implemented in the \texttt{ABINIT} code.\cite{GON02} We have used a plane wave basis set (40 Ry cut-off) with Martins-Troullier norm conserving pseudopotentials. Periodic unit cell large up to 30 Bohr (for Pt) and 35 Bohr (for pDPP) along x and y directions (where z is the direction of the polymer chain) has been necessary to get rid of spurious replica interactions. In particular, the screened Coulomb potential W has been approximated by using the standard Plasmon Pole Model~\cite{GOD89} (PPM). 

In order to converge our GWA calculations, we have made use of 1200 (PT) and 1000 (pDPP) bands for PPM calculations for both W and $\Sigma$. 

The goal of our DFT-B3LYP calculations is to investigate the effects of varying the number of thiophene units within the polymer. Resulting HOCO, LUCO and band gap obtained from DFT are shown in Fig.~\ref{fig2}. The first column deals with pDPP which shows the lowest band gap. The number of thiophene unit in the copolymers rises as one goes from left to right. From this figure, it is clear that increasing the number of thiophene increases the HOCO level and decreases the LUCO level of polymers. Hence, the band gap of polymers increases with the number of thiophenes. Both polymers directly to the right of PCBM, whose synthesis appear possible, show a LUCO level comparable to the PCBM. The three polymers with the highest number of DPP units show a LUCO level even lower.

The two last columns deal with pBBTDPP2 (see Fig. \ref{fig1}b) and PT which as been investigated experimentally.\cite{Bur08, Wie08,Carlos-PT,Sakurai} %
It is noteworthy that the DFT-B3LYP band gaps coincide with the experimental optical gap of pBBTDPP2~\cite{Wie08} and PT~\cite{CHU84}. These coincidences occur with calculations neglecting the e-h interaction energy, giving us confidence that DFT-B3LYP band gaps can be used as prediction for other polymers made with the same building blocks.

\begin{figure}[!b]
\begin{center}
  \includegraphics[scale=.35]{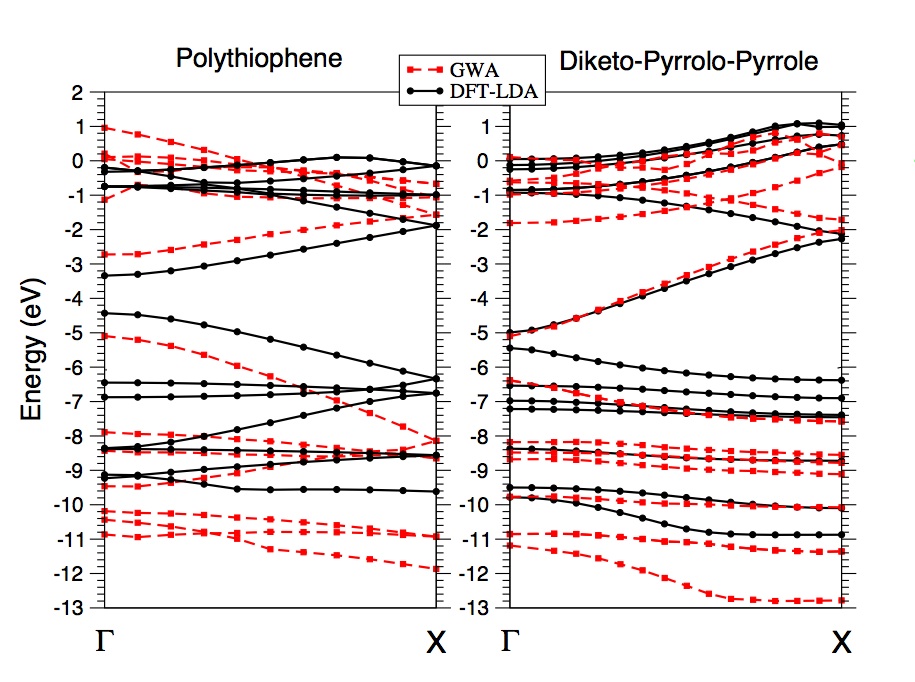}
  \caption{(Color online) Band plot of pDPP and PT: black solid thick: DFT-LDA; red circles and dashed lines: GWA}
  \label{fig3}
  \end{center}
\end{figure}

In what follow, we will show that GW 
results are in good agreement with the B3LYP ones for pDPP and PT, which are at the two extrema of Fig~\ref{fig2}. This allow us to conclude that B3LYP results on intermediate polymers should, in principle, also agree with GW
, supporting further our B3LYP conclusions. In Fig. \ref{fig3}, we have shown the GW (PPM) PT and pDPP band electronic structures. In the PT isolated chain, the 1.1 eV calculated DFT-LDA band gap is increased up to 2.38 eV in GWA
. This is in disagreement with former GWA PT isolated chain calculations (3.58 eV).\cite{J-WvandHorst_PRL,J-WvandHorst_PRB2000,J-WvandHorst_PRB2002} A possible reason for this difference stems from smaller unit cell in the previous results not large enough to isolate the polymer chains in GWA calculations. Nevertheless, their three dimensional approximation where the chain is embedded in a macroscopic dielectric medium is more in agreement to our results (2.49 eV). In GWA , the slope of $\pi$ HOCO valence band is enhanced: starting with negligible GW corrections with respect to DFT-LDA at $\Gamma$ point, at X the energy difference between the two methods is 1.1 eV. Moreover, the intensification of the correlation effects produce the lowering of the core valence bands from 0.2 eV up to 1 eV. In contrast, the GWA corrections result in $\pi^*$ LUCO conduction band almost rigidly shifted. 
In isolated pDPP chain, the almost 0.6 eV DFT-LDA band gap is evaluated 1.284 eV in GWA
. Whereas now the dispersion of HOCO band is almost unmodified, the GWA corrections for LUCO band vary from ~0.4 eV at $\Gamma$ to 0.8 eV at X point. The valence bands decrease in energy as typical behavior in GW calculations for semiconductors. On the other hand, the GWA conduction bands get more matted and shift down in energy.


In conclusion, we have found a polymer with a theoretical LUCO level comparable to the PCBM one, suggesting the later could potentially play the role of the former in photovoltaic devices. We also found three polymers with LUCO energies even lower, which could eventually be used as electron acceptor. The pDPP shows a great dispersion of the conduction band leading to believe it should be a good charge carrier. GWA calculations for PT show the direct band gaps equal to 2.38 eV (PPM)
.
 In pDPP, the GWA band gap is calculated 1.28 eV
 GW
  results are in good agreement with the B3LYP ones.

\begin{acknowledgments}
We are grateful to Mario Leclerc, Matteo Giantomassi, Fabien Bruneval and Simon Pesant for helpful comments. This work is supported by NSERC and FQRNT. Computational resources were provided by the RQCHP.
 \end{acknowledgments}

\bibliography{apssamp}

\end{document}